\documentclass[twocolumn,english]{IEEEtran}
\usepackage[T1]{fontenc}
\usepackage[a4paper]{geometry}
\usepackage{float}
\usepackage{graphicx}

\makeatletter

\providecommand{\LyX}{L\kern-.1667em\lower.25em\hbox{Y}\kern-.125emX\@}

\let\oldforeign@language\foreign@language
\DeclareRobustCommand{\foreign@language}[1]{%
  \lowercase{\oldforeign@language{#1}}}

\makeatother

\usepackage{babel}

\begin{document}

\title{\textbf{Interferometric Observations of Geosynchronous Satellites}}

\author{C.R. Subrahmanya, Peeyush Prasad, R. Somashekar\\Raman Research
Institute,\\C.V. Raman Avenue, Sadashivnagar,\\Bangalore - 560 080,
India%
\thanks{\texttt{crs@rri.res.in, peeyush@rri.res.in, som@rri.res.in}%
}}
\maketitle
\begin{abstract}
In recent years, a large number of geosynchronous satellites are being
planned to provide augmentation services for enhancing the precision
to global positioning systems, e.g., GPS, in applications such as aircraft
landing. In this paper, we present a scheme for co-locating passive
satellite observational facilities with a radio astronomy facility
to open a new possibility of providing valuable data for radio astronomical
imaging, ionospheric studies and satellite orbit estimation.\end{abstract}

\markboth{Interferometric Observations of GEO Satellites}{Subrahmanya, Prasad,
Somashekar}

\section{Introduction}

\PARstart{I}{n} recent years, an increasing number of geosynchronous
satellites are being planned to cope with the need of providing regional
navigation services or augmentation to global systems like the GPS.
From various announcements made by the Indian Space Research Organization
(ISRO), one can expect at least 9 geosynchronous satellites to be
commissioned by ISRO within the next few years with dual frequency
synchronized payloads \cite{key-1}\cite{key-1-1}. Among these, the
first satellite, GSAT-8, has been launched recently and is likely
to commence regular broadcast of WAAS messages from September 2011,
as part of the GAGAN project \cite{key-1}. GAGAN aims to enhance
the precision achievable by GPS-based systems for assisting air-craft
landing. A second geosynchronous satellite with a GAGAN payload has
been announced for launch during early 2012. Both these broadcast
WAAS messages on the L1 (1575 MHz) and L5 (1176 MHz) carriers. 

These satellites will soon be followed by a series of geosynchronous
satellites with dual frequency (L and S-band) navigation payloads
as part of the Indian Regional Navigation Satellite System (IRNSS)\cite{key-1}.
All these, and perhaps some satellites for digital audio broadcasting,
have a unique advantage of continuous visibility from the Indian subcontinent.
(e.g., nine or more L-band signals from geosynchronous satellites
are visible all the time). This brings an interesting combination
of benefits to radio astronomy and satellite orbit estimate requirements.
Here, passive interferometry is used between signals received at different
locations within the space covered by a synthesis radio telescope
like the Giant Meterwave Radio Telescope (GMRT). We refer the reader
to \cite{key-0} for an example of a satellite observation with such
a co-located facility.

\section{Satellite Interferometry: Basic Geometry and Receivers}

\begin{figure}[tbh]
\center{\includegraphics[scale=0.55]{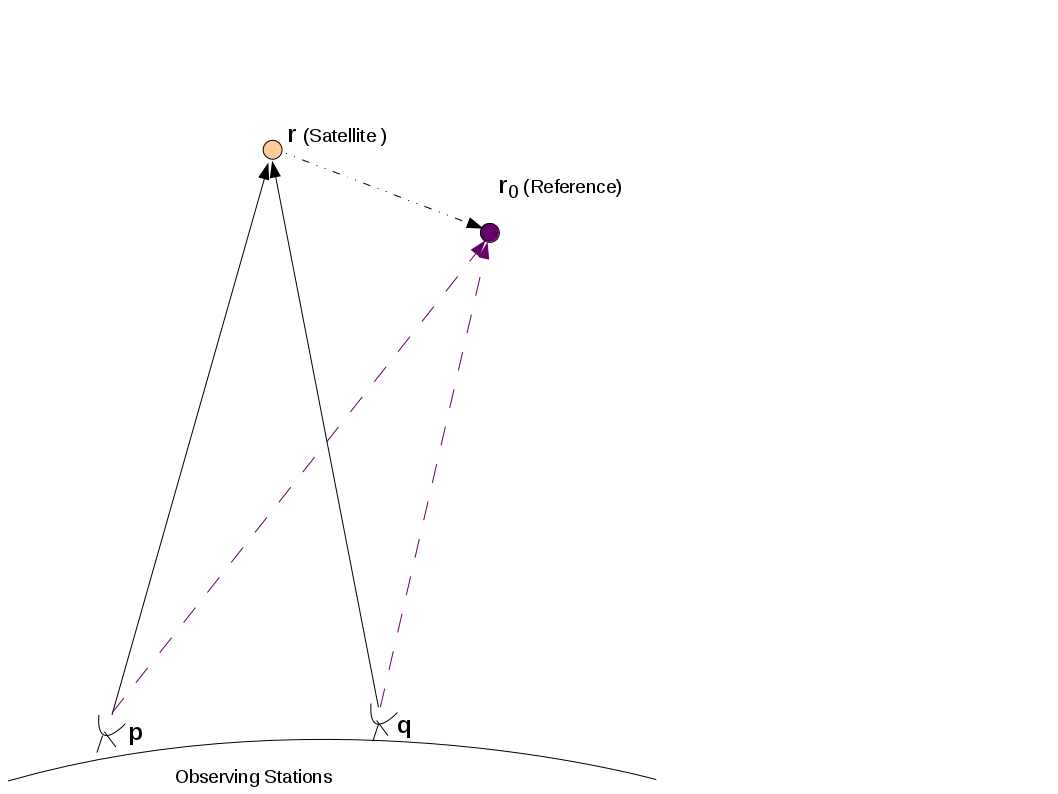}}\caption{\label{fig:SatInt-geometry}Schematic showing the configuration of
the interferometer.}

\end{figure}

A simple illustration of the geometry of interferometry is given in
Fig. \ref{fig:SatInt-geometry}. In this figure, offset of the state
vector with respect to a reference position is represented by $\delta\mathbf{r=}\left(\mathbf{r}-\mathbf{r_{0}}\right)$,
where, $\mathbf{r_{0}}$ is a reference position\textbf{, }and\textbf{$\mathbf{r}$
}is the true position of the satellite. Similarly, satellite velocity
is then given by $\delta\mathbf{v}=\frac{d\left(\delta\mathbf{r}\right)}{dt}$.
A cross-correlation of the signals received at two stations in two
different frequencies can be used to measure instantaneous values
of the components of $\delta\mathbf{r}$ and $\delta\mathbf{v}$ along
the baseline vector joining the two receiving stations. Initial value
for the reference position is generally available (or obtainable as
two-line elements from the Internet) and can be updated using the
interferometric measurements using a suitable orbit propagation software.

\begin{figure*}[tbh]
\center{ \includegraphics[width=16cm]{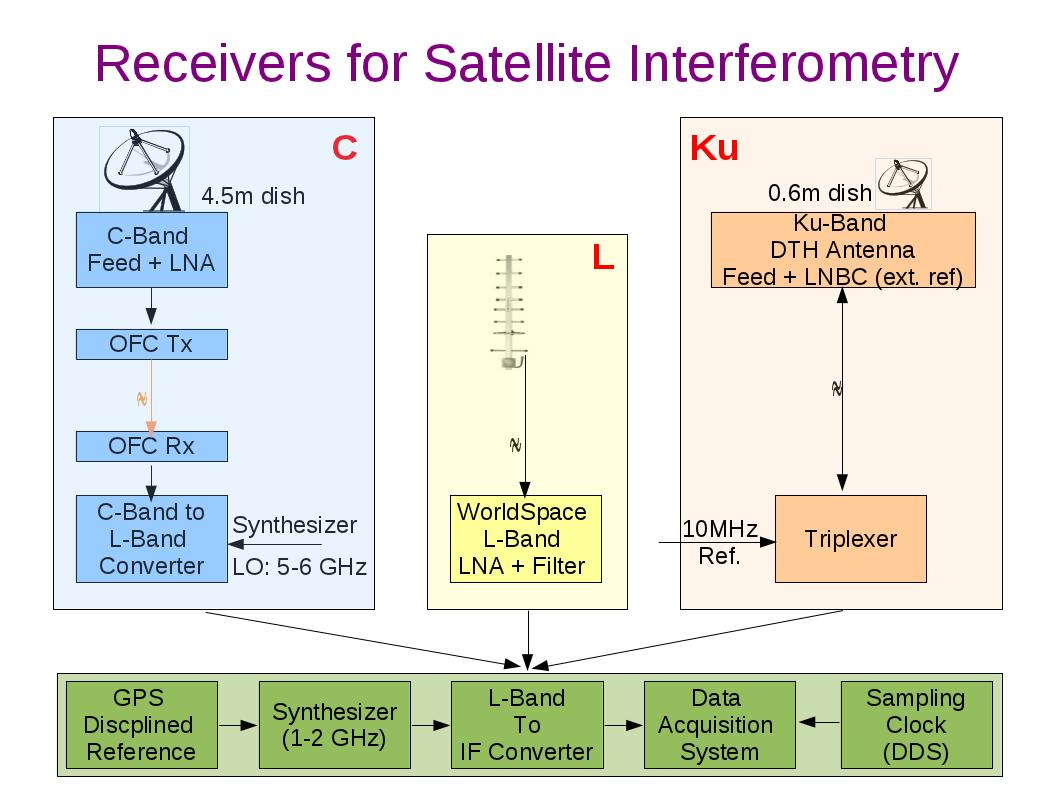}}

\caption{\label{fig:Array-Element-for}Receivers at each array element for
satellite interferometry}

\end{figure*}

In view of the wide variety of inexpensive antennas and receivers
available off-the-shelf for receiving satellite signals, it is possible
to construct a simple system for satellite interferometry (and co-locating
with a Radio Astronomy facility) to obtain valuable data for radio
astronomical imaging, ionospheric studies and satellite orbit estimation.
A preliminary attempt for realizing such a co-located system with
the GMRT was attempted by us in 2005. The subsystems developed for
this project are being augmented to establish a simple facility at
the Raman Research Institute. A conceptual description of the system
constituting this facility has been given in Fig. \ref{fig:Array-Element-for}.

This facility consists of a common L-band subsystem interfaced to
off-the-shelf RF units corresponding to the satellite C, L or Ku-band.
It may be noted that the primary navigation signals are in the L-band,
and provide the most valuable information on the satellite range/Doppler
or ionospheric delay along the line of sight. While this is the most
useful band for measuring ionospheric contribution to interferometric
phases for observing a celestial radio source, the satellite orbit
estimation problem itself requires supplementary data using interferometric
observations which are sensitive to satellite state vector components
along directions perpendicular to the line of sight. In this connection,
we note that the information related to components orthogonal to the
line of sight is a natural outcome of radio interferometry. 

Furthermore, since radio interferometry provides a measure of the
arrival time differences of signals reaching different array elements,
it does not require any knowledge of the nature of signals broadcast
by the satellite. Hence, any band-limited signal transmitted by the
satellite can be used for this purpose. A cross-correlation of the
signals - (after compensating for delay differences resulting from
an assumed reference), provides an estimate of the arrival time differences
in excess of those that can be traced to the reference.

In view of the inherent signal strength coming from satellites, the
required bandwidths and integration times are well within simple processing
capabilities of a normal workstation. However, the angular resolution
which can be achieved by this method (which improves with decreasing
wavelength) depends on the array extent measured in units of wavelength.
We propose to exploit this fact, and the decreasing importance of
ionosphere at higher frequencies, to use the available high frequency
signals from the satellite for interferometry. Interestingly, ALL
the geosynchronous satellites operated by ISRO (including those with
GAGAN or IRNSS payloads) will have their telemetry signals operating
in the C-band. On the other hand, the recently launched satellite
(GSAT-8) with a GAGAN payload has a rich set of communication transponders
in the Ku-band, offering a further advantage for angular resolution.
There are off-the-shelf Ku-band low-noise-block converters (LNBC)
which can take an external reference, and these can be used to provide
coherent translation to L-band for signals from different elements
of the array, by distributing a common local oscillator reference.
Based on these considerations, we have included C and/or Ku-band RF
subsystems to co-exist with a simple L-band antenna (and LNA) for
our array element, as illustrated in Fig. \ref{fig:Array-Element-for}.

\section{Interferometric Observations of Worldspace signals}

As an illustration, we present in this paper some results obtained
in an experiment in satellite interferometry by observing the L-band
signals from a Worldspace satellite (which had a coverage in the Indian
subcontinent during the observations) using commercial Yagi antennas
located at two different GMRT sites. In this experiment, we exploited
the spare capacity of the GMRT fibre-optic network to obtain a long
baseline L-band interferometer. The specific sites chosen for this
experiment were those in the original plan of the GMRT (and hence
with proper termination of fibre-optic cables) but where no GMRT antenna
were installed due to a small reduction of the GMRT antennas arising
from funding constraints. These two locations were used to position
a pair of low-cost antennas (meant for receiving digital audio signals
from Worldspace) effectively separated by about 11.2 kilometres. The
L -band signals received by these low-cost antennas were brought to
a central location using the GMRT fibre-optic network. They were then
down-converted to 70 MHz IF, using a common local oscillator, digitized
and recorded using a PCI data acquisition card. 

\begin{figure}[H]
\includegraphics[width=9cm]{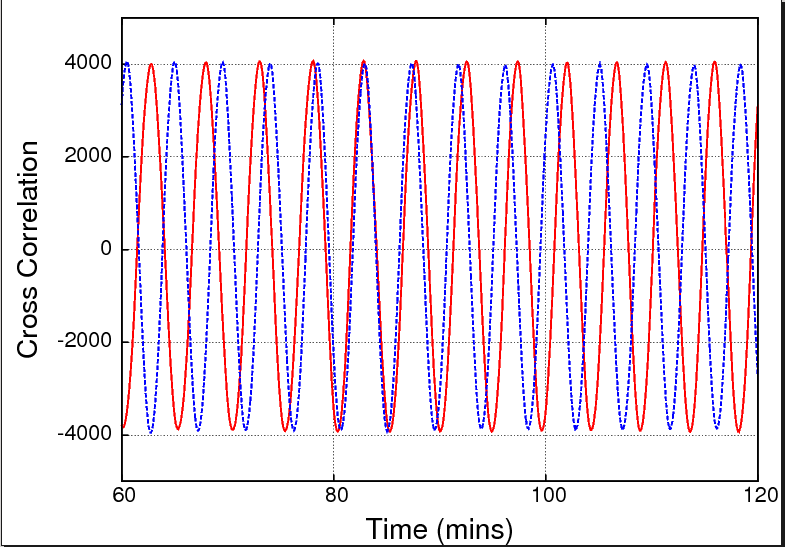}\caption{\label{fig:Plot-showing-the}Plot showing the fringes in two independent
1 hour stretches, separated by an hour.}

\end{figure}

The recorded data were then processed by a simple software to cross-correlate signals
from the two antennas after suitable delay compensation. The slow
relative motion of the satellite with respect to the Earth results
in fringes visible in the estimated cross-correlation, as shown in
Fig. \ref{fig:Plot-showing-the}. The figure shows two sets of fringes
from a one-hour stretch of data, with a one-hour gap between the two
sets. These are typical interferometric fringes, in which a linear
variation of path length difference is translated into a sinusoidal
variation of the cross-correlation. By counting cycles and measuring
the residual phase offset, one can infer that the time taken for a
path length variation of 25 cycles is about 7220 seconds, implying
that the mean time for path length to vary by one wavelength (20.17cm)
was 7220/25 = 288.8 seconds. The signal-to-noise ratio was adequate
for achieving an accuracy of a fraction of degree in phase estimation,
with a one-second integration. Thus, one can conveniently detect a
tiny relative displacement (by a fraction of a millimetre) of the
satellite along a direction parallel to the line joining the two antennas.

Further experiments using the L-band navigation (WAAS) signals and
the Ku-band signals from GSAT-8 are planned soon after the satellite
payloads are officially commissioned for regular services after the
in-orbit-tests. The results will be presented elsewhere.

\section*{Conclusions}

An important conclusion drawn from the above experiments was that
the achievable accuracies for satellite interferometry provide valuable
data for satellite orbit estimation even with much slower phase variations,
as would result from a short (in-campus) baseline. With this in mind,
the facility being established by us is planned to be realized in
a relatively short baseline, within the campus of the Raman Research
Institute.

\section*{Acknowledgements }

Most of the results reported in this paper were obtained from a project
sponsored by ISRO under the RESPOND programme with generous support
from Insat Master Control Facility (MCF). We thank the Director, NCRA
for giving us permission to use the GMRT infrastructure for our experiments
in satellite interferometry. The entire effort has been inter-institutional
and could not have been realized without the whole-hearted support
from many colleagues from the Raman Research Institute as well as
from the GMRT observatory.

\end{document}